\documentclass[12pt]{iopart}
\usepackage{mathrsfs}
\begin{document}

\newcommand\order{\Or}
\newcommand\beq{\begin{equation}}
\newcommand\eeq{\end{equation}}
\newcommand\beqa{\begin{eqnarray}}
\newcommand\eeqa{\end{eqnarray}}
\newcommand\dmu{\partial_\mu}
\newcommand\dnu{\partial_\nu}
\newcommand\dz{\partial_0}
\newcommand\dJ{\partial_j}
\newcommand\dk{\partial_k}
\newcommand\di{\partial_i}
\newcommand\Rmn{R_{\mu\nu}}
\newcommand\Tmn{T_{\mu\nu}}
\newcommand\gmn{g_{\mu\nu}}
\newcommand\hmm{h_\mu^\mu}
\newcommand\hmn{h_{\mu\nu}}
\newcommand\half{\case{1}{2}}

\title[PPN coefficients for Brans-Dicke gravity with $d+1$ dimensions]{Parameterized Post-Newtonian coefficients for Brans-Dicke gravity with $d+1$ dimensions}
\author{Matthew D. Klimek}
\address{Dept. of Physics \& Astronomy, Rutgers University, 136 Frelinghuysen Road, Piscataway, NJ, 08854}
\ead{klimek@physics.rutgers.edu}

\begin{abstract}
We present calculations of Post-Newtonian parameters for Brans-Dicke tensor-scalar gravity in an arbitrary number of compact extra dimensions in both the Jordan and Einstein conformal frames.  We find that the parameter $\gamma$, which measures the amount of spacetime curvature per unit mass, becomes a function of $\omega$, the coefficient of the scalar kinetic term in the Brans-Dicke Lagrangian.  Experiment has placed strong constraints on $\gamma$ which require that $\omega$ become negative in the Jordan frame for any number of extra dimensions, highlighting that this formulation is not physical.  We also confirm the well-known result that a compact extra dimension can be equivalently viewed as a massless scalar `dilaton.'  In the Einstein frame, we find that the behavior of $\gamma$ as constrained by experiment replicates that which is predicted by string theory.
\end{abstract}

\submitto{\CQG}
 
\section{Introduction}
The addition of a scalar field is one of the simplest possible modifications to general relativity.  Tensor-scalar models have been studied as far back as 1948 (Thiry 1948, Jordan 1959).  Theories with a massless scalar were introduced by Brans and Dicke (1961) to satisfy considerations related to Mach's principle.  However, Brans-Dicke gravity continues to be of interest today as string theory also predicts a massless scalar component of gravity (\emph{e.g.} Lidsey \etal 2000).

Another prediction of string theory is compact extra dimensions.  Again, such a notion has been considered for many years in the context of unification, perhaps most notably in the classic papers of Kaluza (1921) and Klein (1926).

It is therefore of considerable interest to study the behavior of Brans-Dicke type theories in an arbitrary number of compact extra dimensions.  The Parametrized Post-Newtonian (PPN) formalism (cf. Will 1993) lends itself to the extraction of observable parameters in a theory-independent manner.  Xu and Ma (2007) calculated the PPN parameters for general relativity with a single compact extra dimension and found that, although theoretically viable, it is ruled out by experiment.  Here we consider Brans-Dicke gravity in an arbitrary number of extra dimensions.  

We complete the calculations in the Jordan frame as the theory was originally formulated.  Nevertheless, several arguments have been made demonstrating that the Jordan frame formulation is in fact unphysical (Faraoni \etal 1999; Magnano and Sokolowski 1994; Cho 1992).  Our results confirm this conclusion.  In the following discussion, we then also transform our results to the physical Einstein frame, and determine the constraints which have been placed on the PPN parameter values by Solar System experiments.

The organization of this paper is as follows. In \S2 we discuss the overall approach to the calculation.  In \S3 we present the actual PPN calculations.  Finally, in \S4 the implications of the PPN coefficients are discussed.

In this paper we follow the signature convention $(- + + + \dots)$.  In \S3 for the sake of brevity, the `=' sign indicates equality to the appropriate order.

\section{Overview of the Calculation}
The Brans-Dicke model postulates a dynamical massless scalar field $\phi$ which couples directly to the Ricci scalar and also includes a kinetic term whose weight is set by the single parameter $\omega$, yielding a Lagrangian density
\beq\mathscr L=R\phi-(\omega/\phi)\dmu\phi\partial^\mu\phi+\mathscr L_\mathrm{matter}.\eeq 
The field equations are obtained by varying (1) with respect to $\phi$ and the metric $\gmn$, and read
\beqa
\nonumber\fl \Rmn-\half\gmn R={8\pi\over \phi}\Tmn+ {\omega\over\phi^2}(\dmu\phi\dnu\phi-\half\gmn(\partial\phi)^2)\\+{1\over\phi}(\nabla_\nu\dmu\phi-\gmn\nabla_\rho\nabla^\rho\phi),
\eeqa
\beq
{2\omega\over\phi}\nabla_\rho\nabla^\rho\phi-{\omega\over\phi^2}(\partial\phi)^2=-R,
\eeq
where $R=\gmn R^{\mu\nu}$ is the Ricci scalar, $(\partial\phi)^2$ is shorthand for $\partial_\mu\phi\partial^\mu\phi$, and $\nabla_\mu$ is the covariant derivative.

We will consider Brans-Dicke theory in $d+1$ dimensions.  Since we know that our universe contains three large spatial dimensions, $d-3$ of the dimensions considered here shall be compact. 
Making use of the fact that in $d+1$ dimensions $\gmn g^{\mu\nu}=d+1$, we may contract the field equations in order to write them in the more useful forms
\beq
\Rmn={8\pi\over\phi}\Big(\Tmn-{1+\omega\over d+(d-1)\omega}\gmn T\Big)+{\omega\over\phi^2}\dmu\phi\dnu\phi+{1\over\phi}\nabla_\nu\dmu\phi,
\eeq
\beq
\nabla_\rho\nabla^\rho\phi={8\pi\over d+(d-1)\omega}T,
\eeq
where again $T=\gmn T^{\mu\nu}$.

Lastly, we will also assume a perfect fluid matter source such that
\beq T^{\mu\nu}=(\rho+\rho\Pi+p)u^\mu u^\nu+pg^{\mu\nu}, \eeq
where $\rho$ is rest mass energy, $\Pi$ is internal (\emph{e.g.} thermal) energy, and $p$ is isotropic pressure.

We wish to solve the field equations for the metric to post-Newtonian order; that is, we want $h_{\mu\nu}$ such that
\beqa g_{00}=-1+h_{00}\sim\order(4)\nonumber\\ g_{0i}=h_{0i}\sim\order(3)\nonumber\\ g_{ij}=1+h_{ij}\sim\order(2)\eeqa
 in terms of the $d$ dimensional potentials defined as 
\[\nabla^2 U=-4\pi\rho,\qquad\nabla^2 V_i=-4\pi\rho v_i, \] 
\[\nabla^2 \Phi_1=-4\pi\rho v^2,\qquad\nabla^2 \Phi_2=-4\pi\rho U, \]
\[\nabla^2 \Phi_3=-4\pi\rho\Pi,\qquad\nabla^2 \Phi_4=-4\pi p, \] where $\nabla^2$ indicates the spatial Laplacian.

In accordance with standard PPN bookkeeping, we count velocity $v\ll1\sim\order(1)$.  Virial relations then give $v^2\sim U\sim p/\rho\sim\Pi\sim\order(2)$. Finally, $\dz=v\cdot\nabla\sim\order(1)$ where $\nabla$ is the spatial gradient operator.

Having obtained the metric to sufficient order, we will compare it to the standard expansion of the metric in terms of the PPN parameters, 
\beqa \nonumber\fl g_{00}=-1+2U-2\beta U^2-2\xi\Phi_W+(2\gamma+2+\alpha_3+\zeta_1-2\xi)\Phi_1\\+2(3\gamma-2\beta+1+\zeta_2+\xi)\Phi_2\nonumber+2(1+\zeta_3)\Phi_3\\+2(3\gamma+3\zeta_4-2\xi)\Phi_4-(\zeta_1-2\xi)\mathcal{A}\nonumber\\
\fl g_{0i}=-\half(4\gamma+3+\alpha_1-\alpha_2+\zeta_1-2\xi)V_i-\half(1+\alpha_2-\zeta_1+2\xi)W_i\nonumber\\
\fl g_{ij}=(1+2\gamma U)\delta_{ij}.\eeqa
This expansion of the metric has been specifically designed to endow these parameters with physical significance.  The parameter $\gamma$ is seen to specify how much space curvature is produced per unit mass, while $\beta$ indicates the strength of nonlinearity in the superposition law.  Both these parameters are equal to one in general relativity, with all remaining parameters equal to zero.  The $\alpha$ parameters measure preferred frame effects; $\zeta$ measures violation of conservation of energy-momentum; and $\xi$ measures preferred location effects.  More details can be found in a reference such as Will (1993).

\section{PPN Coefficients}
\subsection{$\zeta$ to $\order(2)$}
To begin, we must solve the field equation for $\phi$.  We may expand the scalar field around a background value $\phi_0$ as $\phi=\phi_0+\zeta+\order(4)$.  This allows us to write \beq\phi^{-1}=(\phi_0+\zeta)^{-1}=\phi_0^{-1}(1-\zeta/\phi_0+\order(4)).\eeq  The $\order(2)$ term will only be necessary when we are working to $\order(4)$ overall. The only $\order(2)$ term in the energy-momentum tensor is $T^{00}=\rho$ while $g_{00}=-1+\order(2)$ so that $T=-\rho$.  The field equation is now
\beq\nabla^2\zeta=-{8\pi\over d+(d-1)\omega}\rho.\eeq
Since $\dz\sim\order(1)$, The D'Alembertian may be reduced to the usual spatial Laplacian.  The equation may now be solved in terms of the Newtonian potential $U$ as 
\beq \zeta/\phi_0={1\over d-1+(d-2)\omega}GU\eeq
where \beq G\equiv{2\over\phi_0}{d-1+(d-2)\omega\over d+(d-1)\omega}.\eeq

\subsection{$h_{00}$ to $\order(2)$}
To $\order(2)$, (4) becomes
\beq -\half\nabla^2h_{00}={8\pi\over\phi_0}\Big(\rho-{1+\omega\over d+(d-1)\omega}\rho\Big)\eeq
which is solved by
\beq h_{00}=2GU.\eeq
Thus we recognize $G$ as Newton's constant which is now a function of the parameter $\omega$.

\subsection{$h_{ij}$ to $\order(2)$}
Gauge freedom in general relativity allows us to make the transformation $\hmn\to\hmn+\dmu\xi_\nu+\dnu\xi_\mu$ for an arbitrary vector field $\xi_\mu$.  Hence by choosing $\xi_\mu$ to satisfy 
$\dmu\partial^\mu\xi_i=-\partial^\mu h_{\mu i}+\half\di\hmm+\di\zeta/\phi_0$
we may enforce the gauge conditions
\beq\dmu h_i^\mu-\half\di\hmm=\case{1}{\phi_0}\di\zeta.\eeq
This allows us to cast $R_{ij}$ in the form
\beq R_{ij}=-\half\nabla^2h_{ij}+\case{1}{\phi_0}\di\dJ\zeta.\eeq
As $T_{ij}$ is at least $\order(4)$, it may be neglected, and the right hand side of the field equation (4) becomes 
\beq R_{ij}={8\pi\over\phi_0}{1+\omega\over d+(d-1)\omega}\rho\delta_{ij}+\case{1}{\phi_0}\di\dJ\zeta.\eeq
Setting these equal, we obtain
\beq h_{ij}=2GU{1+\omega\over d-1+(d-2)\omega}\delta_{ij}.\eeq
From here we may read off the PPN coefficient
\beq\gamma={1+\omega\over d-1+(d-2)\omega}.\eeq

\subsection{$h_{0j}$ to $\order(3)$}
Imposing one additional gauge condition
\beq\dmu h_0^\mu-\half\dz\hmm=\half\dz h_0^0+\case{1}{\phi_0}\dz\zeta\eeq
allows $R_{0j}$ to be written
\beq R_{0j}=\case{1}{4}\dz\dJ h_0^0+\half\dz(\dmu h_j^\mu-\half\dJ\hmm)-\half\nabla^2h_{0j}+\case{1}{2\phi_0}\dz\dJ\zeta.\eeq
The trace term on the right side of the field equation is now $\order(5)$ and may be neglected leaving
\beq R_{0j}=-{8\pi\over\phi_0}\rho v_j+\case{1}{\phi_0}\dz\dJ\zeta.\eeq
These equations are satisfied if
\beq h_{0j}=-2GV_j{d+(d-1)\omega\over d-1+(d-2)\omega}+\half G\dz\dJ\chi,\eeq
where $V_j$ is the vector potential defined above and $\chi$ is the ``superpotential'' defined by
$\nabla^2\chi=-2U$.
This may be rewritten using the potential $W_j$ from Will (1992) as 
\beq h_{0j}=-{1\over2}{3d+1+(3d-2)\omega\over d-1+(d-2)\omega}GV_j-\half GW_j.\eeq

\subsection{$h_{00}$ to $\order(4)$}
We now return and extend the evaluation of $h_{00}$ to $\order(4)$.  Applying our gauge conditions and previous results to the full expansion of the Ricci tensor yields
\beqa\fl  R_{00}=-\half\nabla^2h_{00}^{(4)}+{G\over d-1+(d-2)\omega}\dz^2U\nonumber
-{2d-3+(2d-4)\omega\over d-1+(d-2)\omega}G^2(\nabla U)^2\\+{2+2\omega\over d-1+(d-2)\omega}G^2U\nabla^2U.\eeqa
Using the normalization condition $ u^\mu u^\nu\gmn=-1$, we can solve for 
\beq u^0u^0={1+u^iu^jg_{ij}\over -g_{00}}=1+2GU+v^2+\order(4).\eeq
Substituting into (6) gives, to the necessary order, 
\beq T_{00}=T^{00}g^2_{00}=\rho(1+\Pi-2GU+v^2).\eeq 
Meanwhile, 
\beq T=T^{\mu\nu}\gmn=-\rho(1+\Pi-dp/\rho).\eeq
Recall that we must now also include the factor $1-\zeta/\phi_0=1-GU/(d-1+(d-2)\omega)$ from (9).
We also must evaluate the last term in (4) carefully.
\beq\nabla_0\dz\zeta=(\dz^2-\Gamma_{00}^\mu\dmu)\zeta=(\dz^2+G\partial^iU\di)\zeta\eeq
Rearranging, we find that
\beqa\fl -\half\nabla^2h_{00} =4\pi G\rho\bigg(1+\Pi-{2d-1+(2d-4)\omega\over d-1+(d-2)\omega}GU\nonumber\\+{d+d\omega\over d-1+(d-2)\omega}{p\over\rho}\bigg)\nonumber
+{8\pi\over\phi_0}\rho v^2+2G^2(\nabla U)^2\\-{1+\omega\over d-1+(d-2)\omega}2G^2U\nabla^2U,\eeqa
which is solved by
\beqa\fl h_{00}=2GU-2G^2U^2+2\bigg({d+(d-1)\omega\over d-1+(d-2)\omega}\bigg)G\Phi_1\nonumber\\
+2\bigg({1+2\omega\over d-1+(d-2)\omega}\bigg)G^2\Phi_2+2G\Phi_3\nonumber\\+2d\bigg({1+\omega\over d-1+(d-2)\omega}\bigg)G\Phi_4,\eeqa
where we make use of the fact that \beq\nabla^2(U^2)=2U\nabla^2U+2(\nabla U)^2.\eeq

\subsection{Results}
We must now dimensionally reduce our results before comparing them with the standard four dimensional form of the PPN metric.  This may be accomplished for any source field by integrating over the compact extra dimensions.  For example, in $n=d-3$ extra dimensions, 
\beq\rho^{(4)}=\int\rho\sqrt{|g'|}\ \rmd^nx\eeq
where $g'$ is the determinant of the $n\times n$ extra dimension sector of the metric.  In the post-Newtonian expansion, this part of the metric will be diagonal so
\beq \sqrt{|g'|}=\bigg(\prod_{i=4}^{3+n}g_{ii}\bigg)^{1/2}=(1+2\gamma U)^{n/2}.\eeq
Using the Green's function $\mathbf G$, the $d+1$ dimensional potential may be written in terms of the four dimensional potential plus an $\order(4)$ correction as follows,
\beqa U(\vec x)&=\int\mathbf{G}(|\vec x-\vec x'|)\rho(\vec x')\ \rmd^3\vec x'\ \rmd^n\vec x' \nonumber\\
&= \int\mathbf{G}(|\vec x-\vec x'|){\rho^{(4)}(\vec x')\over(1+2\gamma U)^{n/2}}\rmd^3\vec x'\nonumber\\
&\approx\int\mathbf{G}(|\vec x-\vec x'|)\rho^{(4)}(\vec x')(1-n\gamma U)\rmd^3\vec x'=U^{(4)}(\vec x)-n\gamma\Phi_2(\vec x).
\eeqa
In fact this is the only correction we need consider.  Since the integrands of other potentials are already $\order(4)$, their corrections will be $\order(6)$ and may be neglected.

Effecting the change of variables in (31), we may now compare our results with the standard PPN metric (8),
to find
\beq\gamma={1+\omega\over d-1+(d-2)\omega},\qquad \beta=1,\qquad \zeta_4={d-3\over 3}\gamma,\eeq
with all others zero.

\section{Discussion}
We first note that there is degeneracy within $\gamma$ between $d$ and $\omega$.  For example, $\gamma=1/2$ in both 5D general relativity (obtained by setting $d=4$ and $\omega\to\infty$ in (36), in agreement with Xu and Ma 2007) and in 4D Brans-Dicke with $\omega=0$.  This is a reflection of the fact that a compact extra dimension can be represented as a scalar `dilaton' field.  A scalar field with $\omega=0$ is exactly what is predicted in a such a situation (Carroll \etal 2002).  In two or more extra dimensions the $\omega$ parameter of the dilaton field becomes a function of the field and thus no longer behaves like a Brans-Dicke scalar (at least in the Jordan frame).

The nonzero value of $\zeta_4$ signals a breakdown of energy conservation in the four dimensional effective theory related to the potential $\Phi_4$, that is, related to pressure.  We may understand this as arising from the fact that pressure is (by construction) isotropically distributed, even over dimensions which we cannot access.  $\zeta_4$ returns to zero when we take $d=3$.  A detection of a loss of pressure energy could then be used to unambiguously determine the true dimensionality of spacetime in Brans-Dicke gravity.  However, given that pressure energy densities are so small compared to mass densities in any feasible experiment, there is little hope that we could measure $\zeta_4$ directly.

It is also important to notice that $v^2$, for example in (30), is a velocity norm over all dimensions, including the compact ones, whereas we can only see velocity components in the large dimensions.  Thus motion in the compact dimensions will appear as an anomalous contribution to $h_{00}$.  This anomaly is eliminated if motion is restricted to the large dimensions.  Such a restriction also returns $\zeta_4$ to zero if pressure arises from microscopic particle velocities.

On the other hand, Solar System experiments have succeeded in placing very strong constraints on the value of $\gamma$ which can in turn constrain the Brans-Dicke parameter $\omega$.

Solar System experiments have set a limit $a\equiv |\gamma-1|<2.3\times10^{-5}$ (Will 2006).  Solving for $\omega$ gives \beq\omega=\frac{2-d+(1-d)a}{d-3+(d-2)a}.\eeq
We see that when $d=3$, $\omega\to\pm\infty$ as $a\to0$.  Given such tight bounds on $a$, we can constrain $\omega>4\times10^4$.  Thus the scalar field is rather unimportant dynamically, if it exists at all.  

In contrast, when $d>3$, $\omega$ is held very close to an order unity negative value of $-(d-2)/(d-3)$.  This conflicts with the original specification by Brans and Dicke (1961) of a positive $\omega$.  Assuming $R=0$ for simplicity, we can easily derive the Hamiltonian of the theory,
\beq\mathscr{H}={\pi^2\phi\over4\omega}+{\omega\over\phi}\nabla\phi\cdot\nabla\phi,\eeq
where the conjugate momentum of the scalar field $\pi=2\omega\dot\phi/\phi$.  We see that if $\omega$ is negative, there is no ground state.  In this way we confirm the well-studied result that the Jordan frame formulation of Brans-Dicke theory is unphysical (Faraoni \etal 1999; Magnano and Sokolowski 1994; Cho 1992).  Even for $d=3$, the presence of linear kinetic terms in the right hand side of (2) signals that the energy is not positive definite.  The Jordan frame equations also fail to correctly specify the dynamics of the spin-2 graviton. 

These irregularities can be clarified by making the usual Einstein frame conformal transformation \beq \tilde\gmn=\phi\gmn,  \eeq which, following a field redefinition $\phi/\phi_0=\exp[(\bar\phi-\bar\phi_0)/\sqrt{\omega+d(d-1)/4}]$, casts the Lagrangian in the usual form of Einstein gravity with a minimally coupled free scalar field $\bar\phi$.  This is the unique transformation which eliminates the problems mentioned in the previous paragraph (Magnano and Sokolowski 1994; Faraoni 1998).  We achieve simplicity in the gravitational terms, however, at the expense of picking up a coupling between the scalar field and matter, thus violating the weak equivalence principle.  

If the Einstein frame is physical, experiments will be sensitive to the transformed metric $\tilde\gmn$.  For our purposes, having already found the Jordan frame metric components to post-Newtonian order, we may simply multiply them by our expression (11) for the scalar field to effect the conformal transformation to appropriate order.  Beginning with $h_{00}$ to lowest order we find
\beqa\fl\tilde g_{00}&=\phi_0\bigg(1+\frac{GU}{d-1+(d-2)\omega}\bigg)(-1+2GU)\nonumber\\ \fl&=\phi_0\bigg(-1+\frac{d-3/2+(d-2)\omega}{d-1+(d-2)\omega}2GU\bigg).\eeqa
Since we know that Newtonian gravity holds to lowest order, we recognize a rescaled gravitational constant \beq\tilde G=\frac{d-3/2+(d-2)\omega}{d-1+(d-2)\omega}G\eeq so that $\tilde h_{00}=2\tilde GU$.  By a similar procedure, we find \beq\tilde h_{ij}=2\tilde GU\delta_{ij}\frac{3/2+\omega}{d-3/2+(d-2)\omega}.\eeq  The last factor in this expression is then recognized as our transformed PPN coefficient $\tilde\gamma$.  

When $d=3$, $\tilde\gamma$ is identically 1, and we can get no information  on $\omega$.  However, defining $\tilde a\equiv |\tilde\gamma-1|$ and solving for $\omega$ gives
\beq\omega=\frac{3-d+(3/2-d)\tilde a}{d-3+(d-2)\tilde a}.\eeq
Thus we find that Solar System experiments constrain $\omega$ to a value of  $-1$ for \emph{any} value of $d>3$ to within $\pm1.2\times10^{-5} n^{-1}$ where $n=d-3$ is the number of extra dimensions.  It is interesting to note that this is the same behavior predicted by string theory.  The gravitational terms of string theory effective actions take the form of Brans-Dicke gravity with $\omega=-1$.  Even after dimensional reduction, the predicted parameter value remains at $-1$ regardless of how many extra dimensions are considered (Lidsey \etal 2000), analogous to our finding that $\omega\approx-1$ for any value of $d>3$.  It is also easy to derive the Einstein frame Hamiltonian and confirm that the theory with $\omega=-1$ is classically stable.

It has been argued that if we consider the Einstein frame to be physical and if the minimal coupling of matter is taken as a fundamental principle, then we ought to effect the conformal transformation on the Lagrangian before adding in a matter term (Magnano and Sokolowski 1994).  In this situation, however, the parameter $\omega$ has no independent influence outside of the transformed field $\bar\phi$, and we can learn nothing about it through the techniques discussed in this paper.

\ack

The author would like to thank the anonymous referees for insightful comments that helped to significantly enhance this paper, Manjul Apratim for interesting discussions and encouragement, and Cerro Tololo Inter-American Observatory for hospitality while this work was completed.

\References

\item[] Brans C and Dicke R H 1961 \PR {\bf 124} 925
\item[] Carroll S M, Geddes J, Hoffman M B and Wald R M 2002 \PR D {\bf 66} 024036
\item[] Cho Y M 1992 \PRL {\bf 68} 3133
\item[] Faraoni V 1998 \PL A {\bf 245} 26
\item[] Faraoni V, Gunzig E and Nardone P 1999 {\it Fundam. Cosmic Phys.} {\bf 20} 121
\item[] Jordan P 1959 \ZP {\bf 157} 112
\item[] Kaluza T 1921 {\it Sitz. Preuss. Akad. Wiss.} {\bf 33} 966
\item[] Klein O 1926 \ZP {\bf 37} 895
\item[] Lidsey J E, Wands D and Copeland E J 2000 {\it Physics Reports} {\bf 331} 343
\item[] Magnano G and Sokolowski L M 1994 \PR D {\bf 50} 5039
\item[] Thiry Y 1948 {\it C. R. Acad. Sci.} {\bf 226} 216
\item[] Will C M 1993 {\it Theory and Experiment in Gravitational Physics} (Cambridge: Cambridge University Press)
\item[] Will C M 2006 {\it Liv. Rev. Rel.} {\bf 9} 2006-3
\item[] Xu P and Ma Y 2007 \PL B {\bf 656} 165

\endrefs

\end{document}